\documentclass[reprint,aip,jap,superscriptaddress,showpacs,nofootinbib,citeautoscrip,floatfix]{revtex4-2}

\usepackage{graphicx}
\usepackage{dcolumn}
\usepackage{bm}
\usepackage{flafter}
\usepackage{float}
\usepackage{lettrine}
\usepackage{amsmath}
\usepackage{amssymb}
\usepackage{color}
\usepackage{epsfig}
\graphicspath{{eps+pdf/}}
\makeatother
\begin{document}

\title {High pressure lattice dynamics study of few layer-$\alpha$-In$_2$Se$_3$}
  
\author{Shiyu Feng}
\affiliation {Department of Materials Science and Engineering, Guangdong Technion-Israel Institute of Technology, Shantou 515063, China}
\affiliation{Department of Materials Science and Engineering, Technion-Israel Institute of Technology, Haifa 3200003, Israel}
\author{Anurag Ghosh}
\affiliation{Department of Materials Science and Engineering, Technion-Israel Institute of Technology, Haifa 3200003, Israel}
\author{Gautham Vijayan}
\affiliation{Department of Materials Science and Engineering, Technion-Israel Institute of Technology, Haifa 3200003, Israel}
\author{Ziyi Xu}
\affiliation{Department of Materials Science and Engineering, Technion-Israel Institute of Technology, Haifa 3200003, Israel}
\author{Qian Zhang}
\affiliation{Department of Materials Science and Engineering, Guangdong Technion-Israel Institute of Technology, Shantou 515063, China}
\author{Elad Koren}
\email{eladk@technion.ac.il}
\affiliation{Department of Materials Science and Engineering, Technion-Israel Institute of Technology, Haifa 3200003, Israel}
\author{Elissaios Stavrou}
\email{elissaios.stavrou@gtiit.edu.cn}
\affiliation{Department of Materials Science and Engineering, Guangdong Technion-Israel Institute of Technology, Shantou 515063, China}
\affiliation{Department of Materials Science and Engineering, Technion-Israel Institute of Technology, Haifa 3200003, Israel}
\affiliation{Guangdong Provincial Key Laboratory of Materials and Technologies for Energy Conversion, Guangdong Technion-Israel Institute of Technology, Shantou 515063, China}

\begin{abstract}
Few-layer $\alpha$-In$_2$Se$_3$  has been studied under pressure using Raman spectroscopy in a diamond anvil cell up to 60 GPa (at room temperature). A combination of AFM and  Raman  was used to estimate the thickness of the specimens. While  few-layer $\alpha$-In$_2$Se$_3$ shows identical  structural evolution with the one of the bulk powder-like form of $\alpha$-In$_2$Se$_3$   ( $\alpha$  $\rightarrow$ $\beta^{'}$ $\rightarrow$ IV ), an abrupt $\beta^{'}$ $\rightarrow$ IV phase transition (at 45 GPa) was observed,   in contrast with the case of the bulk specimen where the two phases coexist over a wide pressure range. This is attributed to the difference in specimens morphology, $i.e.$ single crystal and powder in the case of few-layer and bulk $\alpha$-In$_2$Se$_3$, respectively. This study documents the significance of specimens morphology on the observed pressure-induced phase transitions. The methodology developed in this study for performing high-pressure Raman measurements can be applied to other nanodimensional layered materials. 
\end{abstract}

\maketitle

\section{Introduction}
Two-dimensional materials (2DMs) have emerged as an exciting field of research since the discovery of graphene \cite{Novoselov2005}. These materials have gained considerable interest for applications such as  high-density storage and low-energy-consumption nanoelectronics due to their exceptional optical, mechanical, electrical and ferroelectric properties \cite{Mukherjee2020,Senapati2020,Dutta2021}, surpassing those of traditional ferroelectric materials. In their layered bulk form, 2DMs exhibit strong anisotropic properties due to the strong intralayer covalent bonding as opposed to the weak interlayer van der Waals (vdW) interactions. However, when thinned to monolayer or few-layer levels, their mechanical and electronic properties undergo dramatic changes. For instance, graphene demonstrates ultrahigh carrier mobility absent in graphite \cite{Novoselov2005a}, while monolayer MoS$_2$ transitions from an indirect to a direct bandgap semiconductor \cite{Mak2010}. In the specific case of layered $\alpha$-In$_2$Se$_3$, a III-VI semiconductor, a distinct 3D$\rightarrow$2D crossover was observed, highlighted  by the emergence of  quantum confinement effects \cite{Quereda2016}. Moreover, $\alpha$-In$_2$Se$_3$ exhibits robust spontaneous intercoupled in-plane (IP) and out-of-plane (OOP) ferroelectricity in its monolayer form, arising from the lateral shift of the central Se layer \cite{Ding2017}. This unique property positions it as a potential candidate for ferroelectric memory and optoelectronic devices \cite{Ding2017,Mukherjee2022}.

While high-pressure studies have been conducted on layered materials like graphite \cite{Rao1997} and transition metal dichalcogenides (TMDs) \cite{Nayak2014, Chi2014, Ahmad2025}, these systems are often hundreds of layers thick which do not reach the true 2D limits. In contrast, investigations of few-layer 2DMs under pressure remain few, with only limited studies on materials like monolayer WS$_2$/MoSe$_2$ \cite{Ma2021}. In general, pressure is expected to significantly affect the physicochemical properties of $\alpha$-In$_2$Se$_3$, as evident in other layered compounds \cite{Rao1997,Ahmad2025} under pressure. However, key questions, such as how does pressure affect few-layer $\alpha$-In$_2$Se$_3$ in comparison to its bulk form, remain.  

Motivated by the above, in this study we examine the high-pressure structural behavior of a few-layer In$_2$Se$_3$ specimen in comparison with the bulk form. For this reason, we performed high-pressure Raman spectroscopy study  on a few-layer specimen and compare it with our recently published results for the bulk form \cite{feng2025}. We conclude that while both states show similar structural evolution under pressure, the few-layer specimen intriguingly exhibits an abrupt $\beta^{'}$ to IV transition at 45 GPa, contrasting with the extended phase coexistence observed in the bulk samples. We tentatively attribute the origin of this observation to the different morphology between the two states, single-crystal-like $vs$ powder for the few-layer and the bulk forms of In$_2$Se$_3$, respectively. These differences highlight how material morphology can influence the observed phase transitions under pressure. Finally, a novel  methodology was developed for determining $\alpha$-In$_2$Se$_3$ thickness by combining Raman measurements with Atomic Force microscope  topographical data. 

\section{Materials and Methods}
\subsection{Materials}
Commercially available, HQ graphene Company ($>$99.995$\%$), $\alpha$-In$_2$Se$_3$ single-crystal (SC) specimens were  used for the Raman spectroscopy measurements of this study. A  polyvinyl chloride (PVC) surface protection tape (NITTO SPV224 type, manufactured by ProTapes \& Specialties Company) was used for the physical exfoliation process.  The few-layer specimens were mechanically exfoliated from the bulk SC using the PVC tape and transferred onto a standard 525 $\mu$m-thick silicon (Si) wafer. To enhance the deposition  of large thin flakes, we employed a hot exfoliation technique: the tape with exfoliated crystals was attached on the Si substrate and heated at 120°C for 1 minute prior to separation.

A custom substrate consisted of a 285±13 nm  SiO$_2$ layer thermally grown on top of a 25±5 $\mu$m thick Si disc from Sibranch Microelectronics Company was used for the high-pressure experiments. The use of such a thin substrate necessitates from the need to be compatible with the 40-50 $\mu$m  deep sample chambers used in high pressure experiments, see below. To prevent fracture during processing, the fragile Si wafer was mounted on a glass slide using weak adhesive labels (TANEXO) applied to both right and left sides of the wafer. This stabilization method significantly improved handling reliability while maintaining clean exfoliation surfaces.

\subsection{Atomic Force Microscopy}
The sample morphology was investigated using a Dimension Icon atomic force microscopy (AFM) from Bruker, equipped with a PPP-EFM Pt/Ir coated Si tip from Nanosensors. The spring constant of the tip was approximately 2.8 N/m (ranging from 0.5 to 9.5) and the radius was less than 7 nm. The measurements were conducted in a nitrogen-filled glovebox (with H$_2$O and O$_2$ content of less than 1 ppm) using the tapping mode, where the set point and average speed were configured to 219.4 nm and 10 $\mu$m/s, respectively.

\subsection{Raman Spectroscopy}
Raman studies at ambient condition were measured using a WITec Alpha300R confocal micro-Raman spectrometer using a 532 nm laser for excitation with a 100× objective (NA = 0.9; $\Delta\lambda\approx$ 360 nm, 1800 g/mm$^{-1}$ grating) for laser focusing with a CCD camera (UHTS 300). The laser probing spot dimension was 0.36 $\mu$m. The power of the laser was kept around 1 mW to avoid undesirable sample heating and degradation.
Raman studies under pressure were performed using a custom-made confocal micro-Raman with the 532 nm line of a solid-state laser for excitation in the backscattering geometry. The laser probing spot dimension was 4 $\mu$m. Raman spectra were recorded with a spectral resolution of 2 cm$^{-1}$ using a single-stage spectrograph equipped with a CCD array detector. The laser power on  specimens was kept below 1 mW, to avoid any laser-induced decomposition. Ultra-low-frequency solid-state notch filters allowed us to measure Raman spectra down to 10 cm$^{-1}$ \cite{Hinton2019}. 

\subsection{High pressure studies}
Symmetric diamond anvil cells (DAC) with diamond culets of 300-400 $\mu$m in diameter were used. Between the two diamonds, a high-pressure chamber was constructed using a pre-indented Rhenium gasket with a thickness of 30-40 $\mu$m and a central hole diameter of $\approx$ 120 $\mu$m. A small ruby ball was also loaded inside the sample chamber, to measure pressure using ruby luminescence \cite{Syassen2008}. A few-layer specimen with a thickness of $\approx$ 30 nm, corresponding to $\approx$ 30 In$_2$Se$_3$  layers \cite{Osamura1966}, has been attached to a custom-made Si wafer with a thickness of $\approx$ 25 $\mu$m. Following a backside laser micro-machining procedure (see Fig. S1), the specimen-substrate setup with a $\approx$ 60 $\mu$m diameter was separated from the rest of the Si wafer and loaded inside the chamber. The remaining volume was filled with Neon (Ne) pre-compressed to $\approx$ 2KBar (using a gas loader), acting as the pressure-transmitting medium (PTM). Ne, remains fairly hydrostatic at pressures far above its solidification pressure and up to 30+ GPa \cite{Klotz2009}.

\section{Results}
\subsection{Raman based determination of the thickness of few-layer $\alpha$-In$_2$Se$_3$}
Raman spectroscopy is a powerful tool for determining thickness, and subsequently layer count   in nanomaterials. For example, in graphite, the 2D/G bands intensity ratio abruptly increases with decreasing number of  layers \cite{Liu2013} and optical contrast of exfoliated flakes vary with thickness, providing a quick initial assessment \cite{Ni2007}. Moreover,  in 2H-MoS$_2$, the frequency separation between the E$_{2g}$ and A$_{1g}$ Raman modes increases with increasing number of layers  \cite{Lee2010}. In contrast, current methods for determining the thickness of $\alpha$-In$_2$Se$_3$   rely on qualitative comparisons of overall Raman spectral intensities \cite{Zhou2015}, without establishing a quantitative correlation  between thickness and specific spectroscopic properties. Here, we motivate a novel Raman-based approach to estimate the In$_2$Se$_3$ thickness.

Figure 1 shows the normalized, to the Si (substrate)  Raman mode, Raman spectra of specific test points ((b) and (d)) for In$_2$Se$_3$ samples on a regular Si wafer ((a) and (c)) and the corresponding AFM topographies ((e)-(f)). In Fig. 1(b) and 1(d), the Raman spectra of In$_2$Se$_3$ at ambient pressure can be attributed to the $\alpha$ phase \cite{Lewandowska2001}. According to group theory, the Raman active zone center modes are $\widetilde{\Gamma} {\alpha} $= $5A_1$+$5E$. The three most prominent peaks at 104 cm$^{-1}$ (A(LO+TO) mode),  at 180 cm$^{-1}$ (A(TO) mode), at 193 cm$^{-1}$ (A(TO) mode), and a low intensity peak  at 91 cm$^{-1}$ (E mode) can be observed \cite{Lewandowska2001}. Since the In$_2$Se$_3$ specimens  are typically $<$100 nm thick, the excitation laser penetrates the sample and the characteristic Raman peak of the Si substrate at 521 cm$^{-1}$ (see Fig. S2) was also recorded. Thicker samples (e.g., T1) show weaker Si peaks. After normalizing all spectra to the Si peak intensity, the stronger relative In$_2$Se$_3$ Raman signals correspond to higher thickness (see Figs. 1b,d and Table S1).

\begin{figure}[H]
    \centering
    \includegraphics[width=\linewidth]{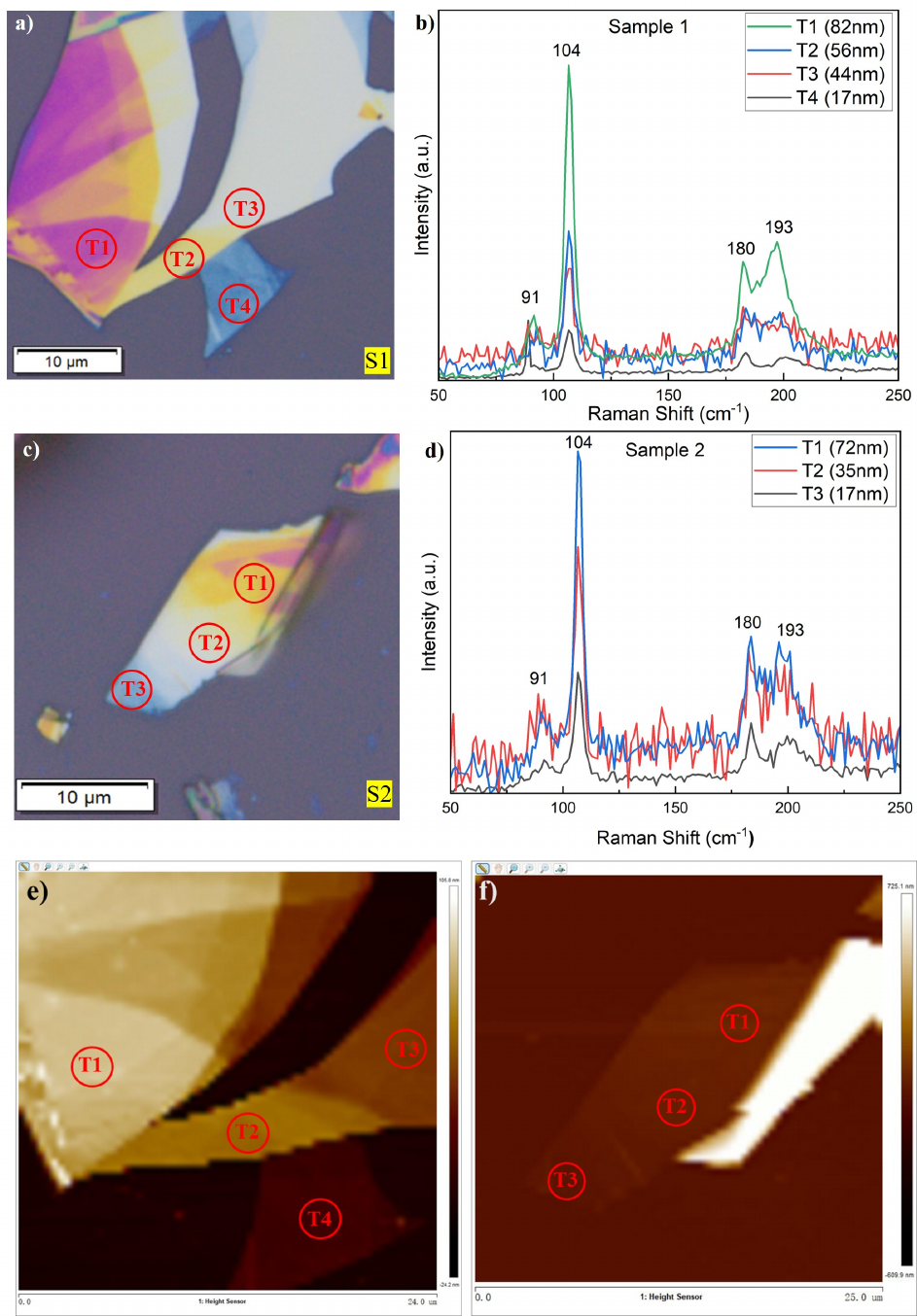}
    \caption{(a) and  (c) Optical microscopy images (100×) of few-layer In$_2$Se$_3$ S$_1$ and S$_2$ specimens exfoliated on  regular Si wafers (525$\mu$m). (b) and (d) Normalized Raman spectra for test points T$_1$-T$_4$ (S$_1$) and T$_1$-T$_3$ (S$_2$). The numbers inside the parentheses are the corresponding thicknesses as determined by AFM. (e) and (f) Corresponding AFM topographies.}
\end{figure}

The intensity ratio I$_{180 cm^{-1}}$/I$_{104 cm^{-1}}$ between the characteristic Raman peaks exhibits a clear thickness dependence, increasing systematically as the specimen thickness decreases (see Table S1). These peaks correspond to distinct vibrational modes \cite{Lewandowska2001}: the 180 cm$^{-1}$ mode arises from in-plane transverse optical (TO) vibrations, which reflect the strength of intralayer covalent/ionic bonding, while the 104 cm$^{-1}$ mode is partially attributed to out-of-plane longitudinal optical (LO) vibrations, sensitive to interlayer van der Waals interactions \cite{Lee2010,Lee2015}. Thus, we tentatively attribute the observed I$_{180 cm^{-1}}$/I$_{104 cm^{-1}}$  trend  to the  effectively  reduced    relative intensity cross-section  of the   interlayer interactions (LO-associated mode at  104 cm$^{-1}$) in comparison  with  the intralayer interactions (TO-associated mode at  180 cm$^{-1}$). Consequently, the   ratio serves as a rough thickness indicator for few-layer In$_2$Se$_3$, with higher ratios corresponding to fewer layers. A similar observation was reported in the case of few-layer  graphene, when the relative intensity of the two E$_{2g}$  modes (intra- and interplanar modes) for different number of layers is compared \cite{Lin2018}. 

Further measurements on 25 $\mu$m-thick  custom-made Si wafers (see Figs. S3-4 and Table S2) show consistent trends: thinner In$_2$Se$_3$ exhibits weaker 104 cm$^{-1}$ and 180 cm$^{-1}$ peaks and a higher I$_{180 cm^{-1}}$/I$_{104 cm^{-1}}$ intensity ratio. For thicknesses in the range of 20-40 nm, the intensity ratios range from 0.45-0.65 (see Fig. 2). This relationship provides a practical guide for identifying a $\approx$30 nm thickness for the In$_2$Se$_3$  flakes used for the high-pressure experiments. The 30 nm specimens ($\approx$30 layers) were selected because they exhibit  detectable Raman spectra, despite the  requirement of  $<$2 mW laser power to avoid decomposition. Thinner samples show almost negligible Raman signals, especially inside the DAC,  and are harder to observe optically due to their transparency. 

\begin{figure}[H]
    \centering
    \includegraphics[width=\linewidth]{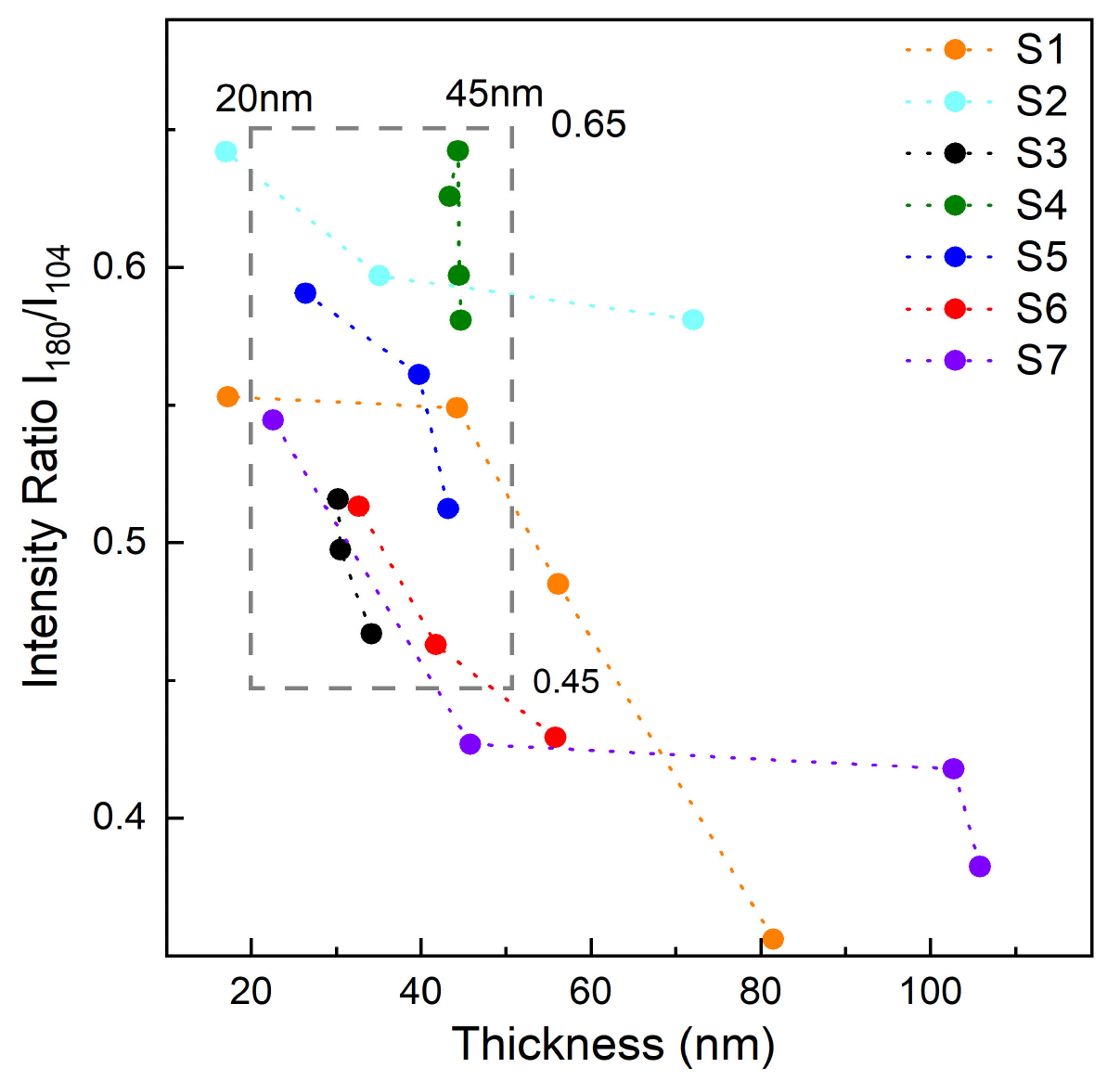}
    \caption{ I$_{180 cm^{-1}}$/I$_{104 cm^{-1}}$ Raman modes  intensity ratio as a function of thickness (as determined by AFM). Data points colors correspond to different probing specimens, S1-S7.}
\end{figure}

\subsection{Raman Spectroscopy of few-layer $\alpha$-In$_2$Se$_3$ under pressure}
Figure S5 shows the Raman spectrum of few-layer $\alpha$-In$_2$Se$_3$ at ambient pressure outside the DAC on top the thin SiO$_2$/Si substrate. The I$_{180 cm^{-1}}$/I$_{104 cm^{-1}}$  intensity ratio  was determined to $\approx$ 0.49, indicating a  $\approx$30 nm thickness.  Selected high-pressure Raman spectra and the corresponding Raman mode frequencies as a function of pressure of the few-layer $\alpha$-In$_2$Se$_3$ are presented in Figs. 3 and 4, respectively. These plots, after comparison with the ones of the bulk $\alpha$-In$_2$Se$_3$ under pressure, reveal that the general structural evolution of the few-layer $\alpha$-In$_2$Se$_3$ is consistent with that of the bulk $\alpha$-In$_2$Se$_3$ \cite{feng2025}. In details, Raman spectra can be assigned to  the $\alpha$ phase up to 1.6 GPa. Above this pressure, In$_2$Se$_3$  transforms to the $\beta^{'}$ phase that remains stable up to 45 GPa.  Finally, In$_2$Se$_3$   transforms to the IV phase, based on the appearance of a new mode at $\approx$ 258 cm$^{-1}$ and the apparent increase of the Raman mode  frequency $vs$ pressure slopes above this pressure, as was also observed in the bulk  In$_2$Se$_3$  at around the same pressure \cite{feng2025}. However, in the case of the few-layer In$_2$Se$_3$ state this transition appears abrupt in contrast to the case of the bulk state that the two phases coexist over a wide pressure range. We postpone the detailed discussion about this critical difference for the discussion section of our manuscript. 

\begin{figure}[H]
    \centering
    \includegraphics[width=\linewidth]{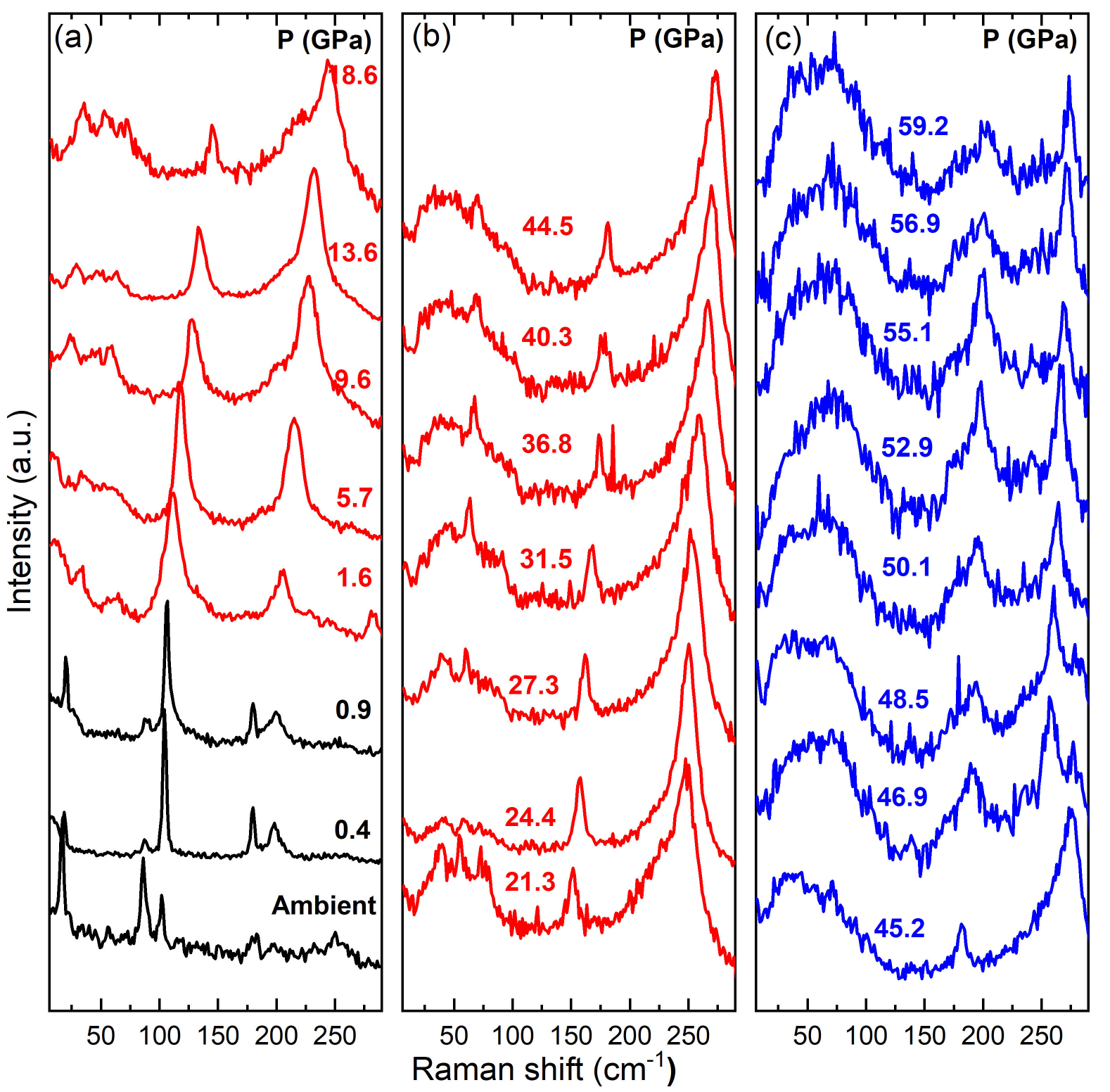}
    \caption {(a)-(c) Selected high pressure Raman scattering spectra at different pressure ranges for few-layer In$_2$Se$_3$.  We note that the spectrum at ambient conditions was acquired outside the DAC, where Raman signals show strong orientation dependence \cite{Ferraro2008}. In contrast, the spectrum at 0.4 GPa was measured inside the DAC, where diamond's high refractive index (n $\approx$ 2.4) scrambles laser polarization, minimizing orientation-dependent intensity variations \cite{Born2017,Gruodis2003}. Consequently, the Raman spectra    resemble the ones \cite{feng2025} of  ``powder-like" specimens.}
\end{figure}

In figure 4, a slight mismatch between the Raman mode frequencies of the few-layer and of the bulk In$_2$Se$_3$  can be clearly observed above 30 GPa. This originates from the use of  Ne  as the PTM in this study, due to its lower compressibility than helium, which prevents excessive gasket hole shrinkage and substrate breakage. It becomes slightly  non-hydrostatic, compared with helium that was used as PTM for the bulk In$_2$Se$_3$ measurements, above 30 GPa \cite{Klotz2009} (see Fig. S6). Therefore, a slight shift of the Raman mode frequencies of the few-layer specimen above 30 GPa towards higher pressures will result to  an overlap with the ones of the  bulk specimen, see discussion about the effect of non-hydrostatic conditions on the apparent Raman mode frequencies as a function of pressure in Ref. \cite{feng2025}. 

\begin{figure}[H]
    \centering
    \includegraphics[width=\linewidth]{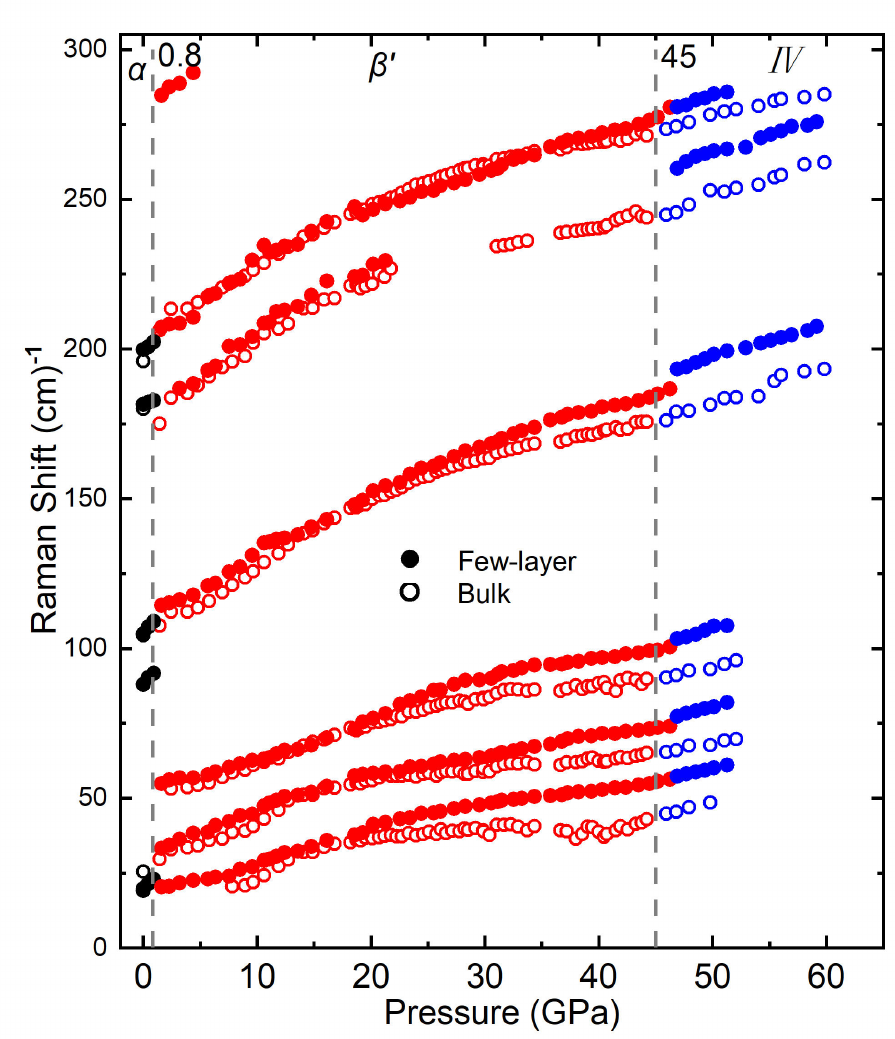}
    \caption {Pressure dependence of the In$_2$Se$_3$ Raman mode frequencies. The frequencies of the few-layer (solid circles) and bulk (open circles) In$_2$Se$_3$ and $\alpha$, $\beta^{'}$, and IV phases are shown in black, red and blue respectively. The vertical dashed lines indicate the critical pressures of the phase transitions. }
\end{figure}

It is worth noting that, during  compression, the  Si substrate appears darker than In$_2$Se$_3$  at low pressures but becomes high-reflective above 12 GPa as shown in Fig. S7 (a) and (b). The characteristic Si Raman peak at 521 cm$^{-1}$ although  visible below 12 GPa, disappears above this pressure, see Fig. S7 (c). Both the aforementioned observations are in agreement   with the previously reported metalization of Si under pressure and its effect on the Raman spectra of Si \cite{Guo2010}.

During decompression (Figs. S8-S9), few-layer In$_2$Se$_3$ transforms back to the $\alpha$ phase with minimal hysteresis, showing reversible transitions consistent with the one of the bulk specimen \cite{feng2025,vilaplana2018,zhao2014}. The recovered $\alpha$ phase retains all characteristic Raman modes \cite{Lewandowska2001}, though with weaker 91 cm$^{-1}$ and 104 cm$^{-1}$ peaks as shown in Fig. S5. Broader peaks suggest residual pressure-induced disorder and sample fragmentation upon release. Moreover, we note a broad Raman feature in the spectrum of the released specimen at $\approx$225-275 cm$^{-1}$, that cannot be explained at this stage.   

\section{Discussions}

The Raman spectra of the bulk state \cite{feng2025} and the few-layer state of In$_2$Se$_3$ above 44 GPa are plotted and compared  in Fig. 5. Ultimately, both the few-layer and the bulk In$_2$Se$_3$ transition to the same high-pressure phase IV above $\approx$45-50 GPa. The few-layer $\beta^{'}$ and IV phases exhibit similar frequency $vs$ pressure slopes as the bulk state, indicating identical interatomic bonding in both states. However, certain spectroscopic differences can be identified after the transition to the IV phase, based on the Raman spectra of Fig. 5. First, the main spectroscopic signature of the IV phase is the appearance of a new Raman mode (at $\approx$ 242 and 258 cm$^{-1}$ for the bulk and the few-layer specimen, respectively)  just below  the higher frequency mode of the $\beta^{'}$ phase \cite{feng2025}.  The slight blue shift, in comparison to the bulk specimen, of the IV phase peak of the few-layer specimen  arise from slightly non-hydrostatic conditions caused by Ne compared to helium, as discussed in the results section. 

\begin{figure}[h]
    \centering
    \includegraphics[width=\linewidth]{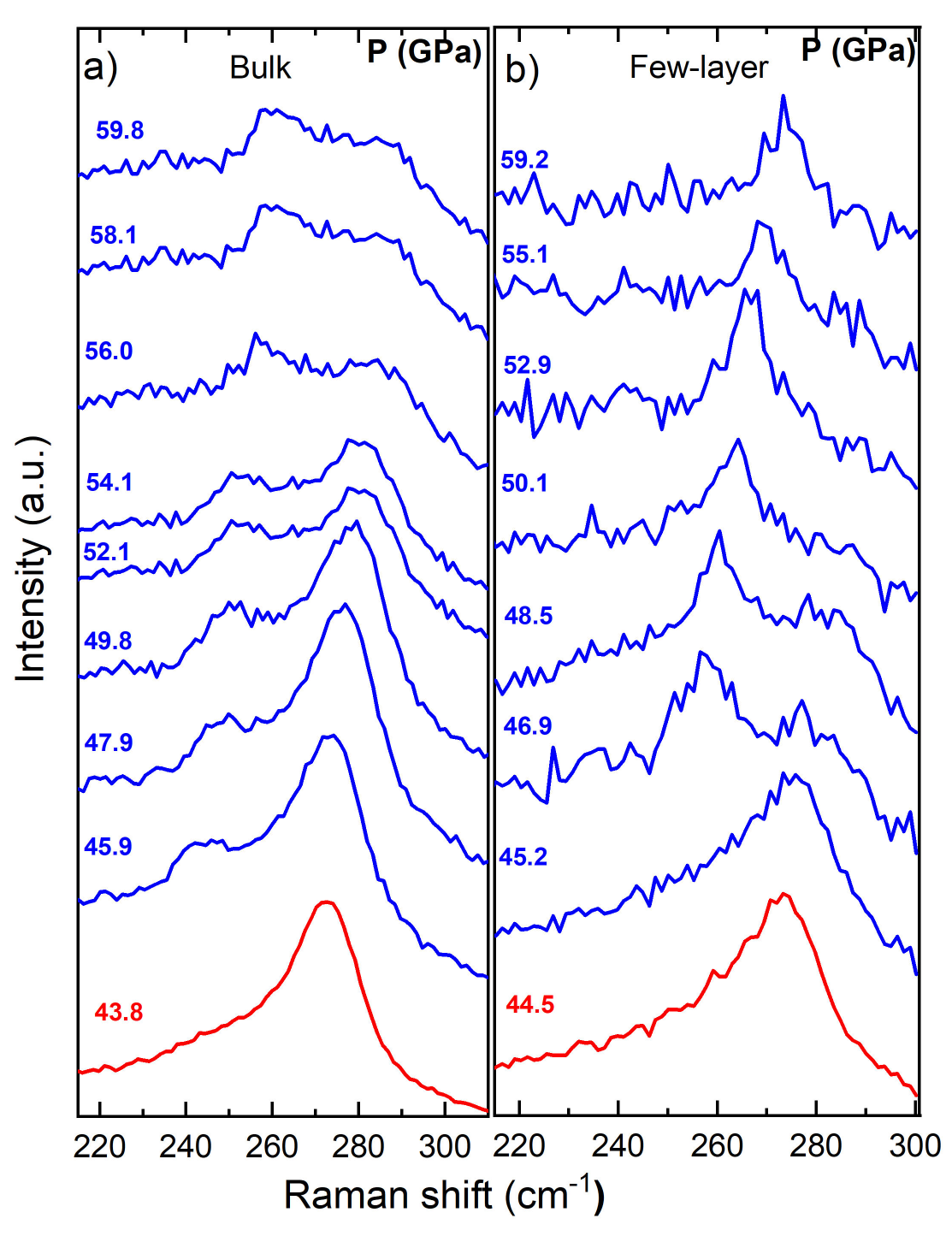}
    \caption {Selected high pressure Raman scattering spectra of a) bulk and b) few-layer In$_2$Se$_3$ from 43.8 GPa to 59.2 GPa. The $\beta^{'}$ and IV phases are shown in red and blue, respectively.}
\end{figure}

However, the main spectroscopic difference is the relatively abrupt change of the relative intensities of the two peaks corresponding to the IV (lower frequency) and $\beta^{'}$ (higher frequency)  phases in the case of the few-layer state. This in turns implies an abrupt   $\beta^{'}$ $\rightarrow$ IV transition in the case of the few-layer state as opposed to the gradual transition, $i.e.$ coexistence of the two phases over an extended pressure range \cite{feng2025}, in the case of the bulk state. This difference can be explained on the basis of the different morphology of the two specimens.  The specimen  in bulk state is in powder-like form, composed of many small, randomly oriented crystallites. The presence of these randomly oriented crystallites, together with possible deviatoric  distribution, can result to a broader range of pressures over which  phases can  coexist. In other words,   distinct grains might have different phases, during the compression, when the enthalpy differences between the two phases before and after the transition are relatively small. This explains the gradual intensity inversion between the 242 cm$^{-1}$ and 272 cm$^{-1}$ peaks (see Fig. 5(a)), consistent with XRD results \cite{feng2025}. 

On the other hand, the specimen in few-layer form is SC-like, precluding coexistence of phases. Thus, the few-layer In$_2$Se$_3$ transforms abruptly under pressure due to its SC nature, showing minimal phase coexistence in Fig. 5(b). This can be explained by a self-seeding nucleation effect, $i.e.$ even a miniscule part of the SC that undergoes a phase transition can act as a self-nucleation center for the whole SC. This is somewhat similar with the previously observed contact-induced phase transition, when a metastable phase comes into contact with the most stable form of a given compound \cite{Millar2009}.  We note that a similar difference of the phase coexistence pressure ranges between powder (extended)  $vs$ SC (sharp) states was previously observed in the case of the structural evolution of MoS$_2$ under pressure \cite{Goncharov2020}. 

This trend is clear in the pressure dependence of the relative intensity  of two peaks that reflects the relative abundance of the two phases, see Fig. 6. The relative abundance is calculated as each phase's peak intensity divided by their sum, assuming equal Raman cross-sections of the corresponding Raman modes of the two phases due to identical bonding. As it is apparent from Fig. 6, in the case of the few-layer state the transition from $\beta^{'}$ to IV phase occurs within just 2 GPa, reaching $>$70$\%$ completion.  While XRD could confirm the exact structural evolution of the few-layer In$_2$Se$_3$, obtaining signals from  $\approx$30 nm SCs remains challenging due to substrate interference, alignment sensitivity, and peak broadening \cite{Fridrichova2016,Tweedie2018}.

\begin{figure}[h]
    \centering
    \includegraphics[width=\linewidth]{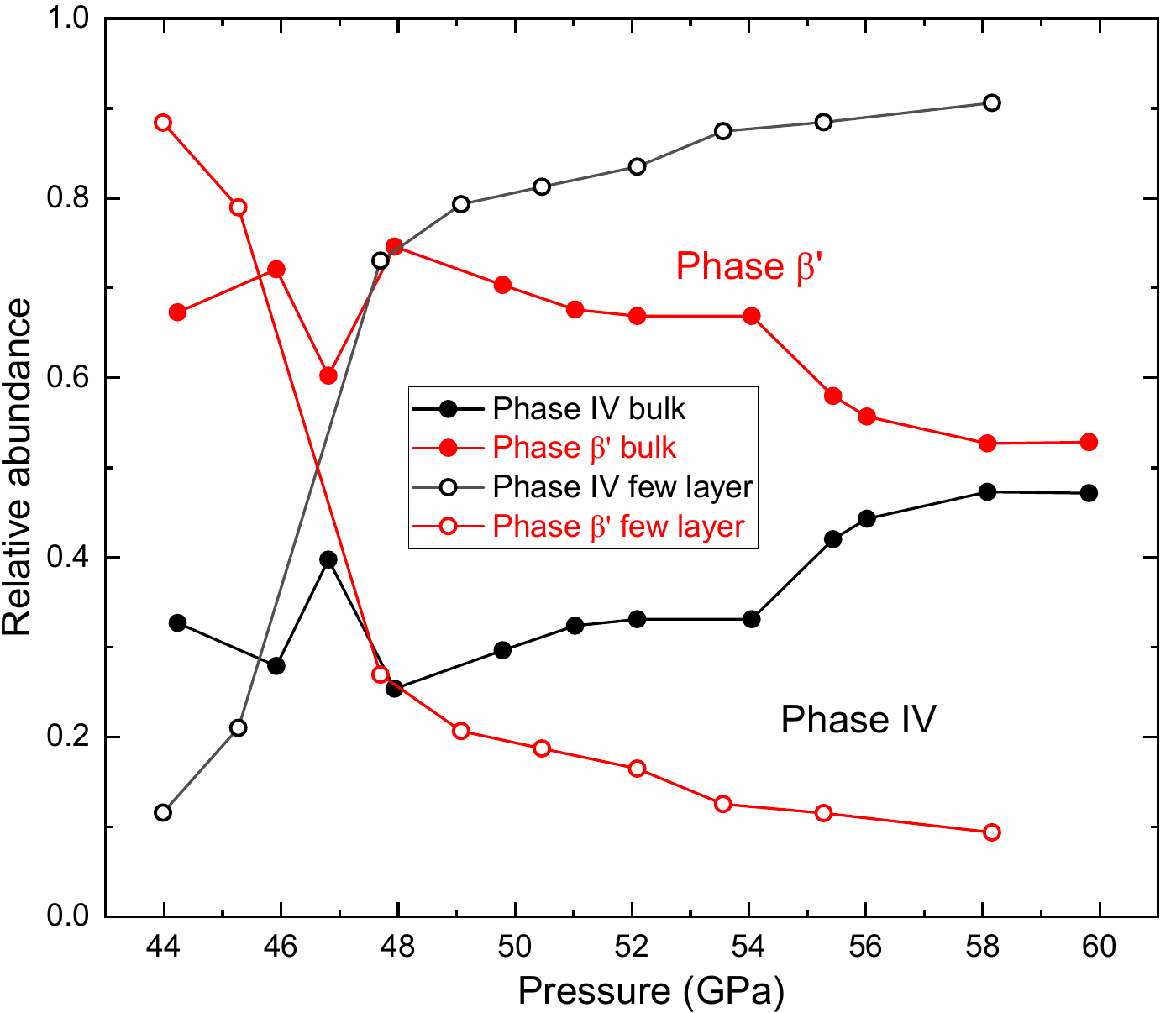}
    \caption {The relative abundance of the $\beta^{'}$ (red symbols) and IV (black symbol) phases, for bulk (solid symbols) and few-layer (open symbols) In$_2$Se$_3$ as a function of pressure, from 43.8 to 59.2 GPa. }
\end{figure}

Finally, we would like to underscore that relevant HP structural studies should be aware of the differences that may arise from the powder $vs$ SC form of the specimens. A phase transition that appears to involve co-existence of phases (both low- and high-pressure phases present for a wide range of pressure) in the powder form of a specimen, will appear to be abrupt in the SC form, especially when the relative enthalpy difference is low. Although in our case the minuscule dimensions of the SC specimens make use of XRD practically impossible, nowadays SCs of the order of nanometers can be readily made for a plethora of materials.

\section{Conclusion}
A novel approach on using Raman spectroscopy to estimate the thickness of few-layer In$_2$Se$_3$ was developed. The intensity ratio between the two most prominent peaks A(TO) mode at 180 cm$^{-1}$ and A(LO+TO) mode at 104 cm$^{-1}$  increases as thickness decreases.  From the comparison between the Raman spectra of the bulk state and few-layer state of In$_2$Se$_3$, the general phase transitions of the few-layer state In$_2$Se$_3$ agree with the result derived from the bulk state. However, above 44 GPa, an abrupt inversion of relative intensities of the higher-frequency  Raman modes was observed for the few-layer specimen, which is associated with minimal  coexistence of the two phases. The experimental methodology for preparing and performing  Raman spectroscopy measurements of few-layer In$_2$Se$_3$ under pressure, can be the base for relevant measurements in other 2DMs under pressure.

\section{Supplementary Material} 
a) Details about the Backside micro-machining procedure

b) Supplementary figures:
Fig. S1: Illustration of the backside micromachining procedure.
Fig. S2: Raman spectra of  test points T$_1$  and T$_2$  of the S$_1$ sample on top a regular Si wafer.
Fig. S3: Microscopy images  and corresponding AFM topographies  of  specimens  S$_3$ and S$_4$.
Fig. S4: Microscopy images  and corresponding AFM topographies  of f specimens  S$_5$-S$_7$ .
Fig. S5: Raman spectra  measured at ambient pressure before compression  and after full pressure release.
Fig. S6: Pressure standard deviation $\sigma$ of 4:1 meth-eth., Silicone oil, He and Ne PTMs as a function of pressure.
Fig. S7: Few-layer In$_2$Se$_3$ with its substrate  placed in the sample chamber. 
Fig. S8: Selected high pressure Raman  spectra upon pressure release.
Fig. S9: Pressure dependence of the  Raman-active mode frequencies upon pressure increase  and release.

c) Supplementary Tables
Table SI: Samples positions, Raman intensity   and thickness  of  specimens S$_1$ and S$_2$.
Table SII: Samples positions, Raman intensity  and thickness  of specimens S$_3$-S$_7$ .

\begin{acknowledgements}
S.F. acknowledges support from the Graduate scholarships of the Guangdong Provincial Key Laboratory of Materials and Technologies for Energy Conversion. E.K. gratefully acknowledges the Israeli Ministry of Innovation, Science \& Technology for financial support.
The work performed at GTIIT was supported by funding from the Guangdong Technion Israel Institute of Technology and the Guangdong Provincial Key Laboratory of Materials and Technologies for Energy Conversion, MATEC (No. MATEC2022KF001).
\end{acknowledgements}

\section*{Data availability}
The data that support the findings of this study are available from the corresponding author upon reasonable request.

\section*{Declaration of competing interest}
The authors have no conflicts to disclose.

\end{document}